\documentclass[aps,prl,twocolumn,superscriptaddress]{revtex4}

\usepackage{epsf,epsfig,amsmath,amssymb,amsfonts}
\usepackage{color}
\usepackage{slashed}

\newsavebox{\ns}
\newsavebox{\dbrane}

\def\be{\begin{equation}}
\def\ee{\end{equation}}
\def\bea{\begin{eqnarray}}
\def\eea{\end{eqnarray}}

\def\Dslash{\,\,{\raise.15ex\hbox{/}\mkern-12mu D}}
\def\Dbarslash{\,\,{\raise.15ex\hbox{/}\mkern-12mu {\bar D}}}
\def\delslash{\,\,{\raise.15ex\hbox{/}\mkern-9mu \partial}}
\def\delbarslash{\,\,{\raise.15ex\hbox{/}\mkern-9mu {\bar\partial}}}
\def\pslash{\,\,{\raise.15ex\hbox{/}\mkern-9mu p}}
\def\calDslash{\,\,{\raise.15ex\hbox{/}\mkern-12mu {\cal D}}}

\newcommand{\vol}{\mbox{vol}}

\newcommand{\nn}{\nonumber \\}

\def\G{\Gamma}

\def\G{\Gamma}

\def\e{\epsilon}

\def\e{\epsilon}

\allowdisplaybreaks

\begin{document}

\title{Supergravity dual of $c$-extremization}

 \author{Parinya Karndumri}
 \affiliation{Department of Physics, Faculty of Science, Chulalongkorn University, Bangkok 10330, Thailand}
 \affiliation{Thailand Center of Excellence in Physics, CHE, Ministry of Education, Bangkok 10400, Thailand}
 \author{Eoin \'O Colg\'ain}
 \affiliation{Departamento de F\'isica, Universidad de Oviedo, Oviedo 33007, Spain}

\preprint{FPAO-13/03}

\begin{abstract}
Recently a general principle, called $c$-extremization, which
determines the exact R-symmetry of two-dimensional CFTs with
$\mathcal{N} = (0,2)$ supersymmetry was identified. In this note  we
show that the supergravity dual corresponds to the extremization of
the $T$-tensor of $\mathcal{N} =2$ gauged
supergravity in three-dimensions. To support this claim, we
demonstrate that the expected central charge of CFTs
arising from twisted compactifications of four-dimensional $\mathcal{N}=4$ SYM
on Riemann surfaces, whose gravity dual is a reduction of five-dimensional
$U(1)^3$ gauged supergravity, is
recovered in the three-dimensional framework.
\end{abstract}

\maketitle

\setcounter{equation}{0}

\section{Introduction} \label{Introduction}
As arguably the most concrete example of the holographic principle \cite{holprinciple}, the AdS/CFT correspondence \cite{AdSCFT} states that any solution of string theory with an Anti-de Sitter (AdS) factor should be equivalent to a conformal field theory (CFT) in one space-time dimension lower. This correspondence and its generalisations have proved instrumental in offering unrivaled insights into the non-perturbative regime of quantum field theories and the quantum nature of black holes. 

Within this context, backgrounds with $AdS_3$ factors are particularly appealing since, in contrast to higher-dimensions, the conformal group in two-dimensions is infinite dimensional and as a result the CFTs are much more tractable. Indeed, it is a well-known fact \cite{Strominger:1996sh} that the entropy of a class of five-dimensional black holes can be derived from the central charge of $\mathcal{N} = (4,4)$ CFTs dual to $AdS_3 \times S^3 \times CY_2$ backgrounds of type IIB supergravity. 

Moreover, what makes three-dimensions historically well-suited to holography is the pioneering, pre-AdS/CFT observation \cite{Brown:1986nw} which states that any consistent theory of quantum gravity in three-dimensions with AdS asymptotics defines a two-dimensional CFT. It is in this spirit that we have witnessed a resurgence in variants of General Relativity, notably Topologically Massive Gravity (TMG) \cite{Deser:1982vy} and New Massive Gravity \cite{Bergshoeff:2009hq}. This interest extends to various (warped) $AdS_3$ black holes \cite{Li:2008dq, Anninos:2008fx}, solutions which also crop up in three-dimensional supergravity \cite{Colgain:2010rg, Detournay:2012dz}, and the microscopic degrees of freedom of the dual field theory. Remarkably, it has recently been suggested that TMG may act as a conduit to holography in asymptotically flat space-times \cite{Bagchi:2012yk}. 

In this letter, working directly with  three-dimensional gauged supergravity, without recourse to higher-dimensional string theory constructions, we show how the exact R-symmetry and central charge of  $AdS_3$ vacua dual to $\mathcal{N} = (0,2)$ CFTs may be identified. As such, our prescription provides a supergravity dual for $c$-extremization \cite{Benini:2012cz,Benini:2013cda}, a recently identified lower-dimensional counterpart of $a$-maximization \cite{Intriligator:2003jj}. Since the R-symmetry can mix with flavour symmetries for supersymmetric theories flowing to IR fixed-points, these respective principles extremise polynomials constructed from 't Hooft anomalies, which are recognised invariances of RG flows, to determine the exact R-symmetry. 

Via AdS/CFT,  $a$-maximization  \cite{Intriligator:2003jj} can be recast in terms of volume minimization of Sasaki-Einstein manifolds \cite{Martelli:2005tp,Martelli:2006yb} so that the Reeb vector dual to the R-symmetry is picked out from a linear
combination of candidate $U(1)$ isometries. Subsequent studies \cite{Butti:2005vn,Eager:2010yu} have shown $a$-maximization and volume minimization to be formally equivalent. More generally, $a$-maximization has an interpretation in terms of
the minimization of the Killing prepotential of $\mathcal{N} =2$
gauged supergravity in five-dimensions \cite{Tachikawa:2005tq}, a fact put to use in \cite{Szepietowski:2012tb} to identify the R-symmetry for a family of supergravity solutions
\cite{Bah:2011vv,Bah:2012dg} based on wrapped M5-branes. 

Recent developments beg the question what is the holographic dual
description for $c$-extremization. To address this problem, we retrace the arguments of \cite{Tachikawa:2005tq} in the natural language of
three-dimensional gauged supergravity and,  in the so-called
$T$-tensor of $\mathcal{N} =2$ gauged supergravity, we identify a
function that, when extremised, determines the R-symmetry and central charge. As we shall see, when the $SO(2)_R \sim U(1)_R$ R-symmetry is gauged, the scalar potential is only a function of $T$, meaning that the extremization of $T$ naturally leads to $AdS_3$ vacua. 

\section{Review of $c$-extremization}
In a non-conformal $\mathcal{N} = (0,2)$ supersymmetric theory with
$U(1)_R$ R-symmetry, the R-symmetry is not uniquely defined and
mixing of $U(1)_R$ with the other Abelian flavour symmetries is
permitted. At a conformal fixed-point this changes and an exact
superconformal R-symmetry is picked out. To identify this exact
R-symmetry at the superconformal fixed-point,
\cite{Benini:2012cz,Benini:2013cda} introduced a ``trial R-current"
\be \Omega_{\mu}^{\textrm{tr}}(t) = J^r_{\mu} + \sum_{M (\neq r)}
t_M J^M_{\mu},
\ee
where $J^r_{\mu}$ is a choice of the R-symmetry
current and $J_{\mu}^M ~(M \neq r)$ are Abelian flavour symmetry
currents. From $\Omega^{\textrm{tr}}_{\mu}(t)$ one constructs a
quadratic function $c_R^{\textrm{tr}} (t)$ which is proportional to
the 't Hooft anomaly of $\Omega^{\textrm{tr}}_{\mu} (t)$: \be
\label{trialc} c^{\textrm{tr}}_{R}(t) = 3 \left(k^{rr} + 2 \sum_{M
(\neq r)} t_M k^{rM} + \sum_{M,N (\neq r)} t_M t_N k^{MN} \right),
\ee where $k^{MN}$ are the 't Hooft anomaly coefficients. Recall
that these anomalies arise in the context of theories with $U(1)^P$
global symmetry when the theory is coupled to non-dynamical vector
fields $A^{M}_{\mu}, M=1,\dots, P, $ in a curved background with
metric $g_{\mu \nu}$. The anomalous violations of current
conservation are then given by
\be \nabla^{\mu} J^M_{\mu} = \sum_{N}
\frac{k^{MN}}{8 \pi} F^{N}_{\mu \nu} \e^{\mu \nu}, ~~\nabla_{\mu}
T^{\mu \nu} = \frac{k}{96 \pi} g^{\nu \alpha} \e^{\mu \rho}
\partial_{\mu} \partial_{\beta} \G^{\beta}_{\alpha \rho}, \nonumber
\ee
where $F^{M} = d A^M$, $T_{\mu \nu}$ is the stress tensor and
$\G^{\beta}_{\alpha \rho}$ is the Levi-Civita connection for $g_{\mu
\nu}$.

The trial $c$-function (\ref{trialc}) can be motivated from a study of
the $\mathcal{N} =2$ superconformal algebra
\cite{Benini:2012cz,Benini:2013cda}. In particular, for supercharges
$\mathcal{Q}$ with R-charge 1, the algebra fixes a relation between
the central charge $c_R$ and the R-symmetry anomaly $c_R = 3
k^{rr}$. In addition, it can be shown in a renormalization scheme
where all currents are primary fields that there are no mixed
anomalies between the superconformal R-current and flavour currents.
This imposes the constraint $k^{r M} = 0, \forall M \neq r$, and
leads to the extremality condition \be \frac{\partial
c_R^{\textrm{tr}}}{\partial t^M} (t_0) = 0, \quad \forall M \neq r.
\ee Since $c_R^{\textrm{tr}}(t)$ is quadratic, there is a unique
solution $t_0$.

\section{$\mathcal{N}=2$ Supergravity}
Here, following the notation of \cite{deWit:2003ja}, we present a
succinct review of  $\mathcal{N}=2$ gauged supergravity in three-dimensions. The field content comprises scalar fields $\phi^i$,
spinor fields $\chi^i$, both with $i =1, \dots, d$, a dreibein
$e^{~a}_{\mu}$, the spin-connecton $\omega^{ab}_{\mu}$ and two
gravitini $\psi^{I}_{\mu}, I =1, 2,$ which transform under the
R-symmetry group $SO(2)_R$.

The target space for scalars is a K\"{a}hler manifold. As such, it is convenient to decompose the $d$ real fields into $d/2$ complex ones and their corresponding complex conjugates, $\phi^i \rightarrow (\phi^i, \bar{\phi}^{\bar{i}} )$. The K\"{a}hler manifold can then be locally written in terms of a metric $g_{i \bar{i}} = \partial_{i} \partial_{\bar{i}} \mathcal{K}$ where $\mathcal{K} (\phi, \bar{\phi})$ is the K\"{a}hler potential.

As explained in \cite{deWit:2003ja}, a subgroup of isometries may be gauged through the introduction of an embedding tensor $\Theta_{MN}$ which defines the Killing vectors that generate the gauge group $X^i = g \Theta_{MN} \Lambda^N(x) X^{Ni}$, where $g$ is the gauge coupling constant and $\Lambda^N(x)$ denotes the gauge group parameters. As is customary, the embedding tensor appears along with gauge fields $A^{M}_{\mu}$ in the definition of covariant derivative
\be
\mathcal{D}_{\mu} \phi^i = \partial_{\mu} \phi^i + g\, \Theta_{MN} A^{M}_{\mu} X^{Ni},
\ee
and also appears in the (Abelian) Chern-Simons (CS) term in the Lagrangian
\be
\mathcal{L}_{CS} = \frac{1}{2} g \e^{\mu \nu \rho} A^M_{\mu} \Theta_{MN} \partial_{\nu} A^N_{\rho}.
\ee
The embedding tensor also crops up in the $T$-tensor $ T = 2 \mathcal{V}^M \Theta_{MN} \mathcal{V}^N$  where $\mathcal{V}$ is the moment map of the gauged isometries. We observe here that the $T$-tensor is quadratic in the moment maps, so structurally it bears some resemblance to the trial $c$-function (\ref{trialc}).

Lastly, the scalar potential of the gauged theory may be expressed
in terms of a real superpotential $F$: \be V = -g^2 \left(8 F^2 -  8
g^{i \bar{i}} \partial_{i} F \partial_{\bar{i}} F \right), \ee where one
can choose $F$ to be one of the eigenvalues of the gravitino mass
matrix $F = - T \pm e^{\mathcal{K}/2} |W|$, where $W$ is the
holomorphic superpotential satisfying $\partial_i \bar{W} =
\partial_{\bar{i}} W = 0$. The potential tells us that, even in
the absence of gauging, one can generate a cosmological constant
with constant $W$. An alternative way to do this involves gauging
the R-symmetry group, in which case $T$ is a non-zero constant with
$W=0$. When the R-symmetry is gauged $W$ must vanish since it
transforms non-trivially under $SO(2)_R$.

\subsection{Dual of $c$-extremization}
Now that we have discussed the rudiments of $\mathcal{N} =2$ gauged supergravity, we can recast the argument of \cite{Tachikawa:2005tq} in terms of three-dimensional language. We start by noting that the embedding tensor $\Theta_{MN}$ encodes the CS terms and as observed in \cite{Benini:2013cda} these correspond to the 't Hooft anomalies $k^{MN}$. The exact relationship for wrapped D3-brane geometries we will introduce later.

Next, we remark that for two-dimensional superconformal theories,
the corresponding $AdS_3$ dual geometry will preserve four
supersymmetries. In particular, one can verify that the Killing
spinor equations \cite{deWit:2003ja} are satisfied when $\partial_i
T = 0$. Going further, from an analysis of the anti-commutator of the
supercharges acting on the scalars, one can infer that the
superconformal R-symmetry is \be \label{Rsymmetry} R = \tilde{s}^M Q_M = t
\mathcal{V}^M Q_M, \ee where $Q_M, M =1, \dots, P,$ are charges
corresponding to the currents $J^M_{\mu}$ and $t$ is a constant of
proportionality. As in \cite{Tachikawa:2005tq}, the gauge
transformation for the gravitino \cite{deWit:2003ja} \be
\mathcal{D}_{\mu} \psi^{I}_{\nu} = \partial_{\mu} \psi^I_{\nu} + g
\Theta_{MN} A^M_{\mu} \mathcal{V}^{NIJ} \psi^J_{\nu} \cdots \ee
allows us to use the fact that the gravitino has R-charge one to fix
the constant of proportionality, i.e. $\tilde{s}^M \Theta_{MN}
\mathcal{V}^N =1$,  leading to 
\be \label{s} \tilde{s}^M = 2 \,T^{-1}
{\mathcal{V}^{M}}, 
\ee where $T$ is the $T$-tensor we introduced
earlier. We are now in a position to propose the supergravity trial
$c$-function
\be
\label{thetac}
c_R \propto \tilde{s}^N\Theta_{MN}\tilde{s}^M  = 2 \,T^{-1}.
\ee
Observe that this trial function is extremised when $\partial_i
T = 0$ which is precisely the condition for a supersymmetric $AdS_3$
vacuum. Furthermore, for D3-branes wrapped on a Riemann surface $\Sigma$, we can infer the constant of proportionality from (3.15) of \cite{Benini:2013cda},
\be
\label{ktheta}
k^{MN} = \frac{\eta_{\Sigma} \,d_G}{2} \Theta_{MN},
\ee
where $d_G$ is the dimension of the gauge group $G$ and $\eta_{\Sigma}$ is related to the volume of the Riemann surface $\frac{1}{2 \pi} \vol_{\Sigma} = \eta_{\Sigma}$. We now recall that the trial $c$-function (\ref{trialc}) is of the form $c_R \sim 3 k^{MN} \sim \frac{3}{2} \eta_{\Sigma}\, d_{G} \Theta_{MN}$, where we have used (\ref{ktheta}). This suggests that the trial $c$-function from the supergravity perspective should be
\be
c_{R} = \frac{3 \eta_{\Sigma} \, d_G}{T}.
\ee

In the next section, we show that this formula recovers the expected central charge for the wrapped D3-brane
geometries discussed in \cite{Benini:2012cz,Benini:2013cda}.

\section{$AdS_3$ vacua from D3-branes}
In this section, to back up our claim, we revisit the initial
example of $c$-extremization presented in \cite{Benini:2012cz}
(later in \cite{Benini:2013cda}), but here recast it in the language
of three-dimensional gauged supergravity. Our point of departure
will be five-dimensional $U(1)^3$ gauged supergravity, which in turn
may be embedded into type IIB supergravity in ten dimensions
\cite{Cvetic:1999xp}. The action reads \bea e^{-1} \mathcal{L}_5 &=&
R  - \frac{1}{2} \sum_{i}^2 (\partial \varphi_i)^2  - \frac{1}{4}
\sum_i^3 X_i^{-2} F^i_{\rho \sigma} F^{i \rho \sigma} \nn &+& V_5 +
\frac{1}{4} \e^{\mu \nu \rho \sigma \lambda} F^1_{\mu \nu} F^2_{\rho
\sigma} A^3_{\lambda}, \eea where $e$ is the determinant of the
vielbein, $A^i$ denotes the gauge fields, $V_5$ labels the potential
\be V_5 = 4  \sum_i^3 X_i^{-1} , \ee and for completeness we define
the constrained scalars \be X_1 = e^{-\frac{1}{2} \left(
\frac{2}{\sqrt{6}} \varphi_1 + \sqrt{2} \varphi_2 \right) }, ~~X_2 =
e^{-\frac{1}{2} \left( \frac{2}{\sqrt{6}} \varphi_1 - \sqrt{2}
\varphi_2 \right) }, \ee with $X_3$ following from the constraint
$X_1 X_2 X_3 = 1$. Observe also that for simplicity we have set the
gauge coupling of the $U(1)^3$ theory to unity $g=1$. This theory
permits the following chain of further consistent truncations:
$\{\varphi_2 = 0, F^1 = F^2 \} \rightarrow U(1)^2$ gauging and $ \{
\varphi_1 = \varphi_2 = 0, F^1 = F^2 = F^3 \}  \rightarrow $ minimal
gauged supergravity, where in the latter case the retained gauge
field is the graviphoton.

To establish a connection to three-dimensions, we adopt the
following ansatz for five-dimensional space-time \be ds^2_5 =
e^{-4A} ds^2_3 + e^{2A} ds^2 (\Sigma), \ee where $A$ is a scalar
warp factor and $\Sigma$ is a Riemann surface with constant
curvature $\kappa = -1, 0, 1$. In tandem, we take an appropriate
ansatz for the field strengths \be F^i =  -a_i \vol_{\Sigma} +
G^i, \ee where closure of $F^i$ implies that $a_i$ are constants and
that associated to each $G^i$ we have gauge potential $B^i$, $G^i =
d B^i$. In addition, we make the natural assumption that the scalars
$\varphi_i$ do not depend on the coordinates of the Riemann surface.

Plugging the ansatz into the five-dimensional equations of motion and reconstructing the Lagrangian, or alternatively performing the reduction at the level of the action, one finds a three-dimensional theory of the form
\bea
\label{Einsteinact}
e_3^{-1} \mathcal{L}_3 &=& R  - 6 (\partial A)^2 - \frac{1}{2} \sum_i^2 (\partial \varphi_i)^2 \nn &-& \frac{e^{4A}}{4} \sum_i^3 X_i^{-2}  G^i_{\rho \sigma} G^{i \rho \sigma} + V_3 \\ &-&
\frac{1}{4} \e^{\mu \nu \rho} |\e_{ijk}|  a_i \, B^j_{\mu} \wedge G^k_{\nu \rho}, \nonumber
\eea
where the final line corresponds to the topological Chern-Simons (CS) term and the new potential is
\bea
\label{potential}
V_3 =  \sum_i^3 \left[ 4 \frac{e^{-4A}}{X_i} - \frac{1}{2} \frac{e^{-8A}}{X_i^2} a_i^2\right]  + 2 \kappa e^{-6A}.
\eea

We can now dualise the gauge fields to bring the action to the
canonical form of a non-linear sigma model coupled to gravity
\cite{deWit:2003ja}. To do this, we redefine the field strengths \be
\label{duality} G^i = X^{2}_i e^{-4A} * DY_i, ~~D Y_i= d Y_i  -
\frac{1}{2} | \e_{ijk} | a_j B^k \ee and rewrite the fields $e^{W_i}
= e^{2A} X_i^{-1}$.  This rewriting has the added bonus that the
scalars are then canonically normalised. In performing this action,
the CS terms remain and one can check that varying the gauge fields
leads to the duality relations (\ref{duality}).

The structure of $\mathcal{N} =2$ supergravity is now manifest. In
particular, one can see that the scalar manifold corresponds to the
coset $[ SU(1,1)/U(1)]^3$ where each factor is parametrised by a
complex coordinate \be z_i = e^{W_i} + i Y_i. \ee This is in line
with expectations, since in \cite{Colgain:2010rg} the same coset
appears when ungauged five-dimensional supergravity is reduced on an $S^2$. However,
one important distinction here is that the R-symmetry is gauged so
$W=0$. To make the K\"{a}hler structure of the scalar target space
more explicit, we can introduce a K\"{a}hler potential \be
\mathcal{K} = -\sum_i^3 \log (\Re z_i). \ee Now that we understand
the scalar manifold, it is relatively easy to extract the $T$-tensor
\bea T &=& \sum_i^3\left[  \frac{1}{2}  e^{-W_i} - \frac{1}{4}
e^{\mathcal{K}}  a_i e^{W_i} \right], \eea and check that
it reproduces the expected terms in the potential (\ref{potential}).
The required gauging of the R-symmetry can also be verified from reducing
the Killing spinor equations from five-dimensions.

We can now minimise the potential with the supersymmetry condition $a_1 + a_2 +a_3 = - \kappa$ \cite{Maldacena:2000mw} leading to the general supersymmetric $AdS_3$ vacuum presented in \cite{Benini:2012cz, Benini:2013cda}. 
This is also a critical point of $T$ as
expected for supersymmetric critical points.

In terms of $T$, the $AdS_3$ radius is now $\ell= 1/(2T)$. One can
then determine the central charge by using the standard holographic
prescription \cite{Brown:1986nw,Henningson:1998gx}
\be
c_R = \frac{3
\ell }{~~2 G^{(3)}}. 
\ee
resulting in the expression
\bea 
c_R &=& -12 \eta_{\Sigma} N^2 \frac{ a_1 a_2 a_3}{\Theta},\nn
\Theta &=& a_1^2 + a_2^2 + a_3^2 - 2 (a_1 a_2 + a_1 a_3 + a_2 a_3), 
\eea
which is, as expected, in perfect agreement
with \cite{Benini:2012cz,Benini:2013cda}. As an added bonus, one can also confirm that
the exact superconformal R-symmetry (\ref{Rsymmetry}), (\ref{s}) agrees with \cite{Benini:2013cda}:
\be
T_{R} = \sum_{i=1}^3 \frac{2 a_i( 2 a_i + \kappa)}{\Theta} T_i, 
\ee
where $T_i$ are the generators of the $SO(2)^3$ global symmetry \footnote{Here we have used the explicit form of the moment map $\mathcal{V}_{i} = \frac{1}{4} e^{-W_i}, ~i = 1, 2, 3$}. While the canonical R-symmetry can be identified from higher-dimensions \cite{Kim:2005ez}, we believe this is the first statement purely in three-dimensional supergravity. 

\section{Summary}
In this letter, we have proposed a natural three-dimensional
supergravity description of $c$-extremization for CFTs with
$\mathcal{N} = (0,2)$ supersymmetry. In light of the work of
\cite{Tachikawa:2005tq}, it is not too surprising that the
$T$-tensor is the function being extremised. From the gravity
perspective, it is already understood \cite{Freedman:1999gp}
that the holographic $c$-function should be inversely proportional
to the real superpotential, and for certain three-dimensional flows,
this is the case \cite{Berg:2001ty}.  However, the fact that we also recover the R-symmetry is certainly novel and it means that one can identify the R-symmetry directly in three-dimensional supergravity without recourse to higher-dimensions. The task remains to identify the
gauged supergravities corresponding to the wrapped M5-brane examples
presented in \cite{Benini:2013cda}. We also hope to identify three-dimensional gauged supergravities which arise from dimensional reductions of generic wrapped-brane geometries, such as those discussed in \cite{Gauntlett:2006ux,Figueras:2007cn,Gauntlett:2007ph}. 

We have enjoyed discussion with N. Bobev, M. P. Garc\'{i}a del Moral, D. Rodr\'{i}guez-G\'omez \& H. Samtleben. The work of P. K. is partially supported by Thailand Center of Excellence in Physics through the ThEP/CU/2-RE3/11 project and Chulalongkorn University through Ratchadapisek Sompote Endowment Fund. E. \'O C acknowledges support from the research grant MICINN-09- FPA2012-35043-C02-02.

\end{document}